\documentclass[12pt]{iopart}
\pdfoutput=1
\expandafter\let\csname equation*\endcsname\relax
\expandafter\let\csname endequation*\endcsname\relax
\usepackage{amsmath}
\usepackage{graphicx}
\usepackage{float}
\begin{document}

\title{Folded transit photometry}

\author{Ma. Janelle Manuel and Nathaniel Hermosa}

\address{National Institute of Physics, University of the Philippines, Diliman, Quezon City, Philippines, 1101}
\ead{mgmanuel1@up.edu.ph}
\vspace{10pt}
\begin{indented}
\item[]April 2021
\end{indented}

\begin{abstract}
Transit photometry is perhaps the most successful method for detecting exoplanets to date.  However, a substantial amount of signal processing is needed since the dip in the signal detected, an indication that there is a planet in transit, is minuscule compared to the overall background signal due mainly to its host star. In this paper, we put forth a doable and straightforward method to enhance the signal and reduce noise. We discuss how to achieve higher planetary signals by subtracting equal halves of the host star - a folded detection. This results in a light curve with a double peak-to-peak signal, $2R_p^2/R_s^2$, compared to the usual transit. We derive an expression of the light curve and investigate the effect of two common noises: the white Gaussian background noise and the noise due to the occurrences of sunspots. We show that in both simulation and analytical expression, the folded transit reduces the effective noise by a factor of $1/\sqrt2$. This reduction and the doubling of the signal enables (1) less number of transit measurements to get a definitive transiting planet signal and (2) detection of smaller planetary radii with the usual transit with the same number of transit data. Furthermore, we show that in the presence of multiple sunspots, the estimation of planetary parameters is more accurate.  While our calculations may be very simple, it covers the basic concept of planetary transits. 
\end{abstract}

\vspace{2pc}
\noindent{\it Keywords}: Transit Photometry, Exoplanet Detection 
%
%

\maketitle
%
%

\section{Introduction}
Transit photometry, which deals with measuring and capturing planetary transits, is perhaps the most successful and widely used exoplanet detection method to date: amassing roughly 70\% of the total discoveries \cite{Deeg2018,exo}. Planetary transits occur when a planet passes directly between a star and an observer. As seen from a particular vantage point, the planet appears to move across the face of the star, temporarily blocking some of the light radiated by it \cite{transit}. The light radiated or flux of the star will decrease by an amount proportional to the ratio of the areas of the planet and the star \cite{wright2012exoplanet}—measuring this flux as a function of time produces the transit light curve \cite{forgan2013possibility,winn2014transits}.

Due to its simple nature, transit photometry searches can operate on a massive scale. Both ground-based and space-based observing facilities can simultaneously watch hundreds of thousands of stars, resulting in a higher probability of detecting an exoplanet. Moreover, transit photometry is often used due to the important parameters such as planetary size, orbital size, and even the planetary system architecture \cite{fischer2015exoplanet} obtained from just measuring the light curve. When combined with other techniques such as radial velocity, the density of the planet can already be obtained thus determining the planet's physical structure. Planets that have been studied using these methods are arguably the best-characterized of all known exoplanets \cite{Deeg,charbonneau2006extrasolar}.

As successful it may be as a detection method, transit photometry is not without limitations. A major drawback of transit photometry is its susceptibility to false positives. False positives produce a similar signature caused by a genuine planetary transit event \cite{Deeg2018}. In fact, the rate of false positives for transits observed by the Kepler mission could be as high as 40\% in single-planet systems \cite{FPrate}. A transit-shaped event in a light curve can be caused by several astrophysical phenomena such as intrinsic stellar variability \cite{Morton_2016,Collins_2018} and stellar activity \cite{bruno2021stellar,Santos2018}. Sunspots (or solar spots) are frequently considered as a main source of error because they tend to mimic the effect of planetary transits \cite{defay2001bayesian}. 

The occultation of a transiting planet with a sunspot may lead to an incorrect estimate of the planetary parameters \cite{bruno2021stellar,refId0,sanchis2011starspots,sanchis2013kepler}. If the planet passes in front of a sunspot, the apparent transit depth will temporarily decrease. Also, if their signal cannot be singled out, or during a given single transit the planet crosses multiple sunspots, removing the effect of the occultation might be difficult \cite{bruno2021stellar}. In fact, Silva-Valio et al. \cite{silva2010properties} found that, under the same assumption that all crossed features are sunspots, the transit radius of CoRoT-2b might be overestimated by up to 3\%. 

Detection of false positives and confirmation of a transiting object's planetary nature requires additional transit measurements or follow-up detection techniques \cite{Deeg2018,latham2003spectroscopic,alonso2004strategies,deeg2009ground}. In ground-based surveys for transiting planets, false positives usually outnumber true planetary systems by a significant factor \cite{Torres_2010}. Thus, confirmation can only be possible through a supplementary high-precision radial-velocity observation\cite{Cameron2016,O_Donovan_2006}. For space-based surveys such as NASA’s Kepler mission, there are simply too many candidates that a follow-up becomes excessively expensive in terms of measurement time \cite{Torres_2010,Cameron2016}. 

Aside from astrophysical phenomena, additive Gaussian white noises, such as photon and readout noise \cite{defay2001bayesian,adda2000photometrie}, can tremendously alter the signal of the exoplanet. Often, the amplitude of the noise is larger than both the signal of Earth-like exoplanets and state-of-the-art instrumentation limit precision \cite{oshagh2018noise}. Hence, a substantial amount of signal processing is needed to obtain the drowned planetary signal.

\begin{figure}[ht]
    \centering
    \includegraphics[width=1\textwidth]{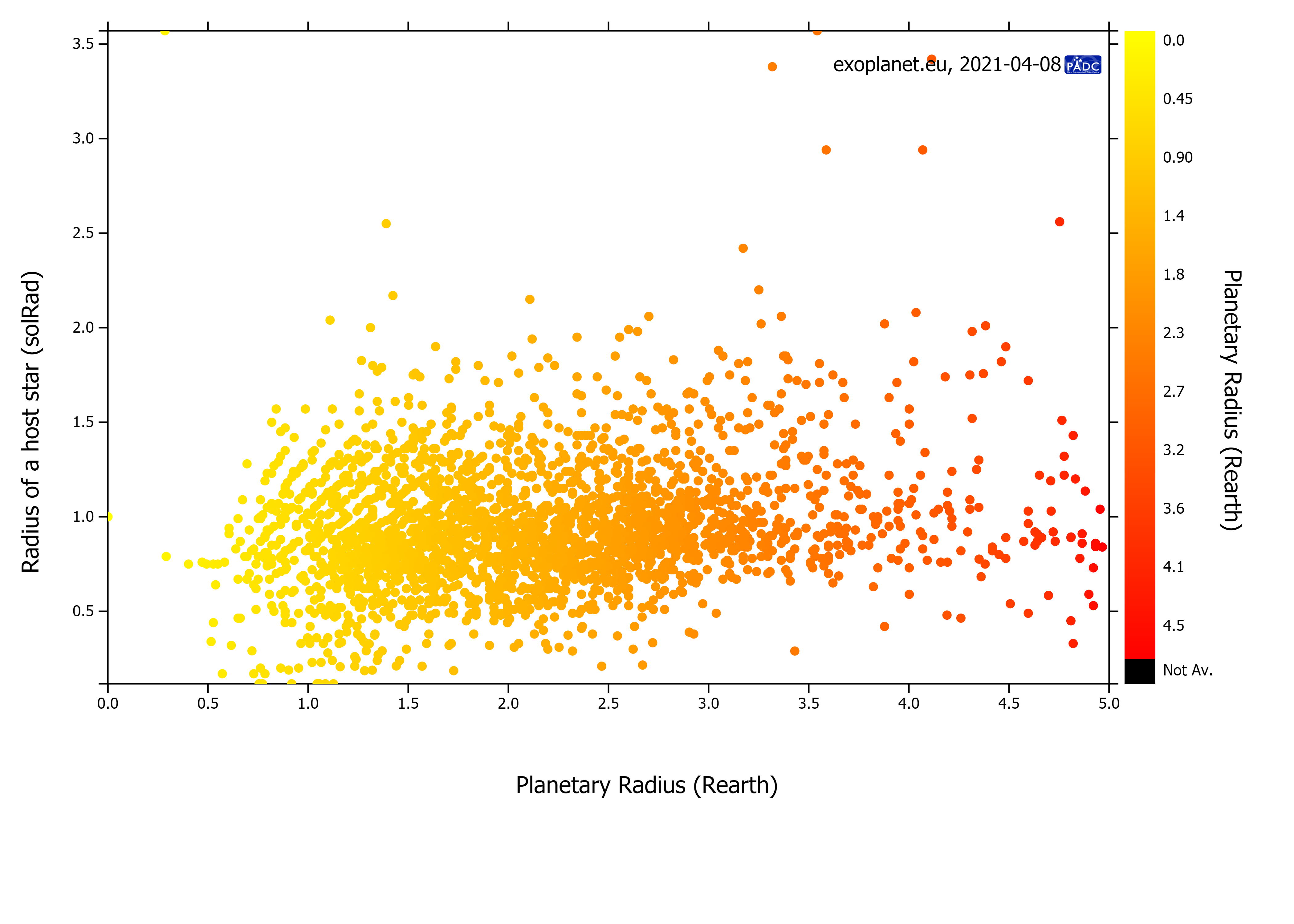}
    \caption{Confirmed exoplanets first detected by transit method as compiled in \cite{exo}. Most exoplanets detected using transit method has host radius between 0.5 to 1.5 of our sun's and a planet radius between 1.0 to 3.0 Earth radius.}
    \label{fig:exo}
\end{figure}

The radius of the host star also plays a vital role in transit photometry. Fig. \ref{fig:exo} shows that most exoplanets detected have a Sun-like star. A massive host star can effectively repress the signal produced by the transiting planet. Planets that are substantially less massive than Earth and Venus or sub-Earths are challenging to detect using this method due to their weak planetary signal \cite{sinukoff2013below}.

This paper presents a way to overcome the drawbacks encountered in the transit method by putting forward a simple technique called folded transit photometry. In folded transit photometry, we increase the planetary signal by subtracting equal halves of the host star. We also explain in detail how to derive the analytical expression describing the folded transit light curve.  In both simulation and analytical expression, we examine how this method reduces background noise and the effect of sunspots on the light curve without trading off the number of transit measurements. Lastly, we show how the Signal-to-Noise Ratio ($SNR$) of this method performs against the existing transit method. 

\section{Derivation of the folded light curve}
In folded detection, we bisect the host star in half with two equal areas, $A_{s,1}$ and $A_{s,2}$.  In our calculations, we proxy the intensity of the host star to be its area. Hence, $\Delta I = I_{2} - I_{1} \propto A_{s,2} - A_{s,1}$.  The folded transit name comes from the subtraction of these two halves. 

\begin{figure}[ht]
    \centering
    \includegraphics[width=0.53\textwidth]{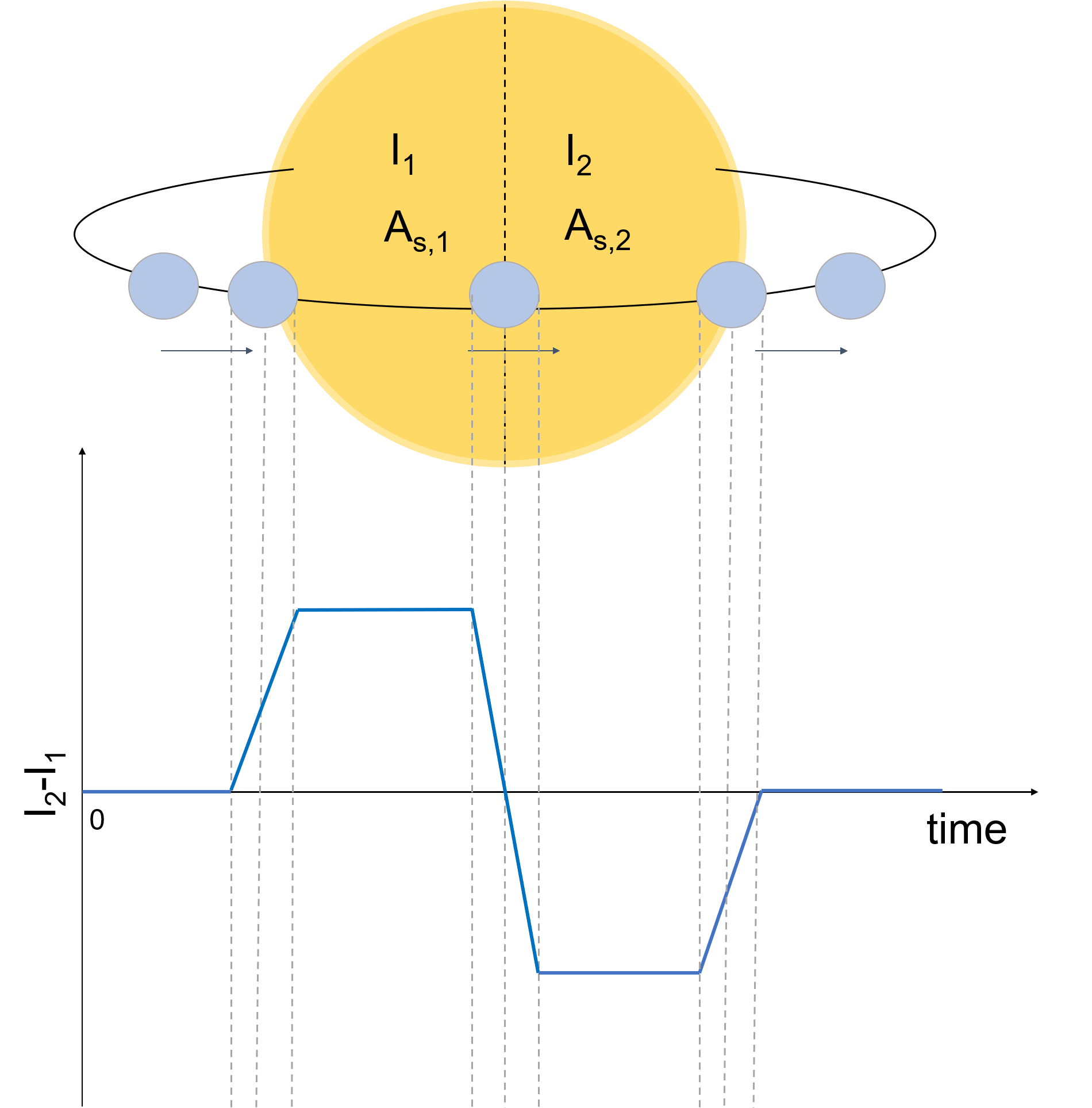}
    \caption{Intuitive depiction of the folded transit light curve. $I_{1}$ is the intensity given by the left half of the host star with area $A_{s,1}$ while $I_{2}$ is the intensity given by the right half of the host star with area $A_{s,2}$. The light curve of the folded transit can be thought of as a signal against a dark background compared to the usual transit curve where the signal is a dip in a high-intensity background.}
    \label{fig:theory1}
\end{figure}

As the planet passes across the host star, we expect a rise and a dip in the $\Delta I$, which corresponds to the planet being in either $A_{s,1}$ and in $A_{s,2}$. Suppose the planet transitions from  $A_{s,1}$ to $A_{s,2}$, the $\Delta I$ will gradually increase at the beginning of the transit and plateaus as the planet moves toward the center. Unlike the usual light curve for transiting planets, the light curve of the folded transit flips its sign when the planet is at the center. After completely crossing the center, the $\Delta I$ plateaus again and increases as the planet ends its transit. This is visualized in Fig. \ref{fig:theory1}.

\begin{figure}[ht]
    \centering
    \includegraphics[width=0.4\textwidth]{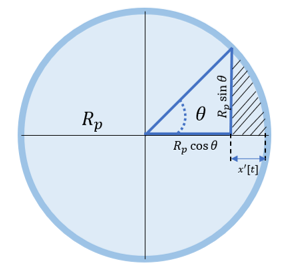}
    \caption{The geometry involved in determining the area of a planet blocking its host star. We assume that $A_{s} >> A_{p}$.}
      \label{fig:theory2}
\end{figure}

The effect of the planet on the $\Delta I$ depends on the area the planet is blocking off from the host star. Therefore, we first have to obtain an expression of the area of the planet, $A_{p}$, as a function of time, $t$. We set up the geometry in Fig. \ref{fig:theory2}.

The hatched area is given by
\begin{equation} \label{eq1}
\begin{split}
A_{p}&= 2 \int  \sqrt{{R_{p}}^{2}-x^{2}} \,dx \\
 & = 2 \int_{0}^{\theta} \sqrt{R_{p}^{2}-(R_{p} \cos \theta)^{2}} R_{p} \sin \theta d\theta,
\end{split}
\end{equation}

where $R_{p}$ is the radius of the planet. The expression inside the integral is just the area of a semicircle.  Evaluating Eq. (\ref{eq1}), we obtain

\begin{equation}
\begin{split}
    A_{p}&=\frac{R_{p}^{2}}{2}\left\{2\theta-\sin(2\theta)\right\}\\
    &= \frac{R_{p}^{2}}{2}\left\{2\cos^{-1}\left(1-\frac{x'[t]}{R_{p}}\right)-\sin\left(2\cos^{-1}\left(1-\frac{x'[t]}{R_{p}}\right)\right)\right\},
\label{areap} 
\end{split}
\end{equation}

where we change $\theta$ in terms of $R_{p}$ and the $x'[t]$ in Eq. (\ref{areap}). This is the equation describing the area of the planet as a function of time. The $x'[t]$ expression is the displacement of the planet, $x'[t]= v \Delta t$, that depends on the region where the planet is currently transiting.

It is evident that there are 5 regions of interest. These are: 1) the moment when the planet starts to enter the host star's area, $0\leq t\leq \frac{2 R_{p}}{v}$; 2) when the planet is within the first half of the host star, $\frac{2 R_{p}}{v}\leq t\leq \frac{R_{s}}{v}$; 3) when the planet is transitioning from the first to the second half, $\frac{R_{s}}{v}\leq t\leq \frac{2R_{p}+R_{S}}{v}$; 4) when the planet is at the second half, $\frac{2R_{p}+R_{s}}{v}\leq t\leq \frac{2R_{s}}{v}$; and 5) the moment when the planet is ending its transit, $\frac{2R_{s}}{v}\leq t\leq \frac{2R_{p}+2R_{s}}{v}$. The time in which the planet is in a certain region depends on $R_p$, $R_s$ (radius of the host star), and $v$ (the velocity of the planet). We derive expressions for all these regions to get the complete light curve as the planet transits.

Determining the $\Delta I$s for regions 2 and 4 are straightforward. The intensities of the halved host star cancels out and we are left with $\Delta I$s that are dependent only on the area of the planet, $A_p$. The difference between these signals, $\Delta I_2 - \Delta I_4 = 2A_p$, is twice the signal that can be detected in the usual transit method. 

In region 3 where the planet moves across the center of the star, the $\Delta I$ is given by

\begin{equation}
\begin{aligned}
    \Delta I_{3}= \frac{{R_{p}}^2}{2}\left\{2 \cos ^{-1}\left[1-\frac{v \left( \frac{R_{s}+2R_{p}}{v}-t\right)}{R_{p}}\right]-\sin \left[2 \cos ^{-1}\left(1-\frac{v \left( \frac{R_{s}+2R_{p}}{v}-t\right)}{R_{p}}\right)\right]\right\} \\ -\frac{{R_{p}}^2}{2} \left\{2 \cos ^{-1}\left(1-\frac{v \left( t-\frac{2R_{s}}{v}\right) }{R_{p}}\right) -\sin \left[2 \cos ^{-1}\left(1-\frac{v \left( t-\frac{2R_{s}}{v}\right)}{R_{p}}\right)\right]\right\},\\
\frac{R_{s}}{v}\leq t\leq\frac{2R_{p}+R_{S}}{v}.
\end{aligned}
\end{equation}

The intensity at the first half increases as the intensity at the second half of the host star decreases. We can determine the velocity of the planet from the slope of $\Delta I_3$ by

\begin{equation}
\begin{aligned}
    m&= -\pi R_{p}v\\
    v&= \frac{-m}{\pi R_{p}}.
\end{aligned}
\end{equation}

The derivation for regions 1 and 5 are quite similar. These are given by

\begin{equation}
\Delta I_{1}=\frac{{R_{p}}^2}{2}\left\{2\cos^{-1}\left(1-\frac{vt}{R_{p}}\right)-\sin\left[2\cos^{-1}\left(1-\frac{vt}{R_{p}}\right)\right] \right\},0\leq t\leq \frac{2 R_{p}}{v}
\end{equation}

\begin{equation}
\begin{aligned}
    \Delta I_{5}=-\pi R^{2}_{p}+\frac{{R_{p}}^2}{2}\left\{2 \cos ^{-1}\left(1-\frac{v \left(t-\frac{2R_{s}}{v}\right)}{R_{p}}\right)-\sin \left[2 \cos ^{-1}\left(1-\frac{v \left(t-\frac{2 R_{s}}{v}\right)}{R_{p}}\right)\right]\right\},\\
    \frac{2R_{s}}{v}\leq t\leq \frac{2R_{p}+2R_{s}}{v}
\end{aligned}
\end{equation}

where the main difference is the $\pi R^{2}_{p}$ term for $\Delta I_5$ because of our construction in Fig. \ref{areap}. 

\begin{figure}[ht]
    \centering
    \includegraphics[width=0.75\linewidth]{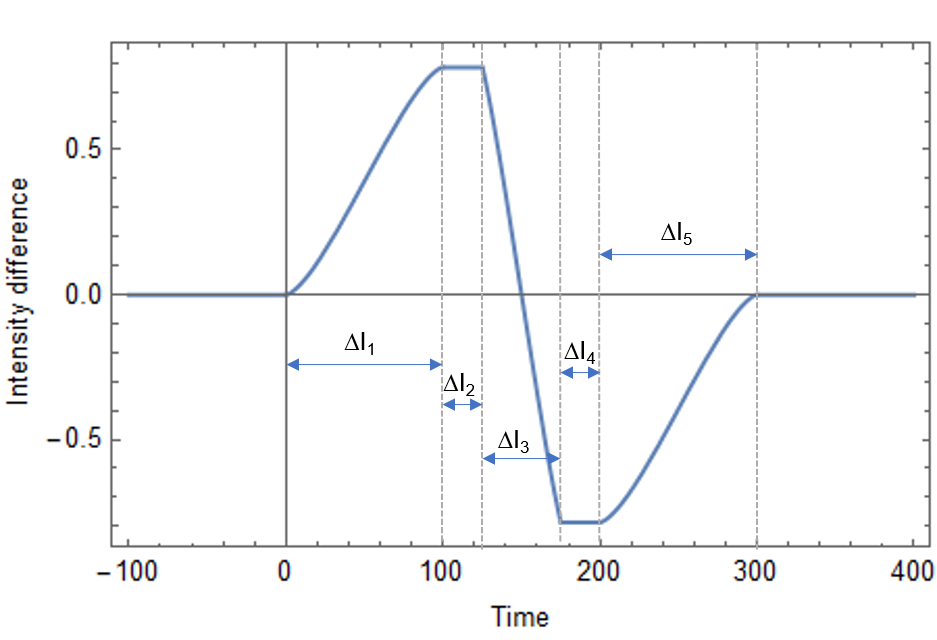}
    \caption{Folded transit light curve derived above. Here the radius of the planet $R_p$ is 0.25 and $R_s$ is 1.} 
    \label{fig:first}
\end{figure}

Finally, we derive the folded transit light curve by summing the $\Delta I$s. The light curve is shown in Fig. \ref{fig:first}. We arbitrarily set the ratio of the radius of the planet to the host star to 1/4 and velocity to 0.01.

The overall width of the light curve is $\text{t}_c=\frac{2(R_{s}+R_{p})}{v}$. Together with the slope $m$, the radius of the planet $R_p$ and its velocity $v$ can be determined, simultaneously. 
\clearpage
\section{Folded transit light curve with noise}
We incorporate two of the most common types of noises encountered in transit photometry\textemdash a white Gaussian signal noise and an external noise due to random sunspots\textemdash in order to have an idea of how the folded transit method performs in actual space conditions. The white Gaussian noise simulates the background noise that may affect the instrument's performance and the natural noise from celestial sources \cite{Deeg,chouhan2015comparative}. On the other hand, the sunspots reduce the intensity of the host star depending on their activity and size.

To add white Gaussian noise to our calculations, we set the host star's intensity to 1 ($R_p^2 = 1$). At any particular time that the signal is detected, the noise changes the host star's intensity value within the variance $\sigma_{bg}^2$. For example, if we have a $\sigma_{bg}=.01$, 68.2\% of all the changes in the intensity of the host star fall within 0.99 to 1.01. In our simulations, we vary $\sigma_{bg}$ from 0.01 to 0.40. When a planet transits, it reduces the intensity of the host star by a fraction of its intensity. A planet that reduces 1\%  of the host star's intensity would mean that the planet's radius is one-tenth of the host star's radius. The calculation is straightforward for the usual transit method, but for the folded transit, half of the signal with its noise is subtracted to the other half of the signal with its corresponding noise. 

We simulate random occurrences of sunspots by setting the circular star host with radius $R_s$ and subtracting areas of random radii with variance $\sigma_{sun}$. The range used for the simulated sunspot radius is consistent with the actual sizes of sunspots observed in our Sun \cite{solanki2003sunspots}. We set the period of each sunspot to be shorter than a single transit measurement time in order to account for various stellar rotation and activity \cite{refId0,silva2010properties,silva2011time}.

\begin{figure}[ht]
    \centering
    \includegraphics[width=0.9\linewidth]{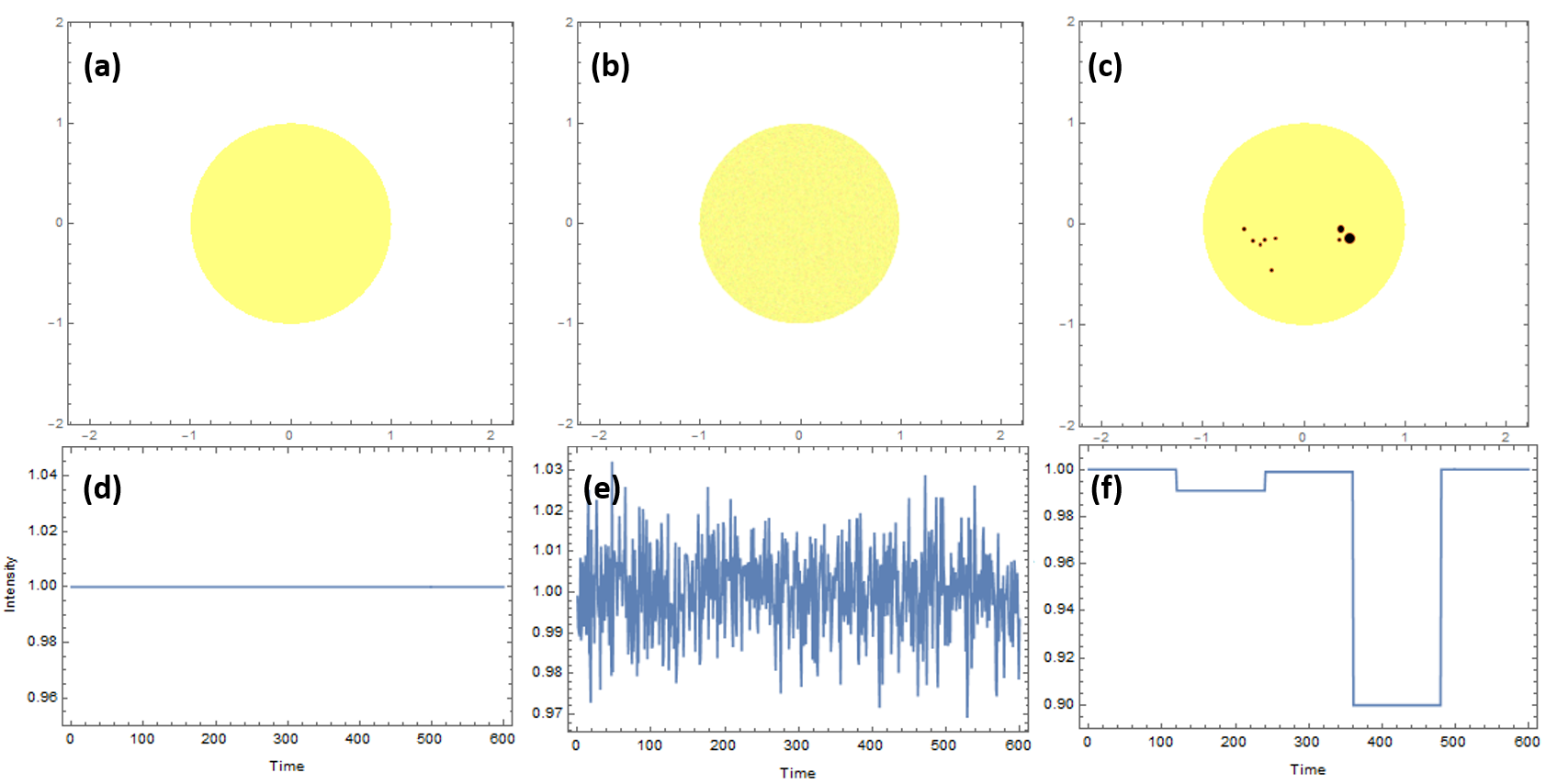}
    \caption{ Simulation of the noise. (a) Illustrations of the host star without any noise, (b) with background noise of $\sigma$ = 0.01 and (c) with sunspots that appear at different times. (d), (e), (f) The resulting light curves with no transiting planet for (a),(b), and (c), respectively. The method to produce these light curves are detailed in the text.}
    \label{fig:simulation}
\end{figure} 

For both the background noise and the random sun spot, we define the output noise $\sigma_{sim}$ as 

\begin{equation}
\sigma_{sim}=\sqrt{\frac{\sum_{i=1}^{N_{tp}}(\Lambda_{n}-\Lambda_{nn})_i^2}{N_{tp}}},
\label{eqn:variance}
\end{equation}

where $N_{tp}$ is the number of on-transit data points as defined by Deeg et al\cite{Deeg}, $\Lambda_n (\Lambda_{nn})$ is the light curve with noise (no noise). $N_{tp}$ depends on the exposure time, the number of transits observed and on the duration of transit. In our case, the duration of transit is the same for both the folded transit and the transit method. 

Light curves for different values of $\sigma_{bg} (\sigma_{sun})$ are calculated and the ratio between the noise from simulation $\sigma_{sim}$ and the input noise $\sigma_{bg}(\sigma_{sun})$ are obtained. Finally, we calculate the $SNR$ for both the transit and the folded transit method.

\subsection{Astrophysical White Noise}
\begin{figure}[ht]
    \centering
    \includegraphics[width=0.55\linewidth]{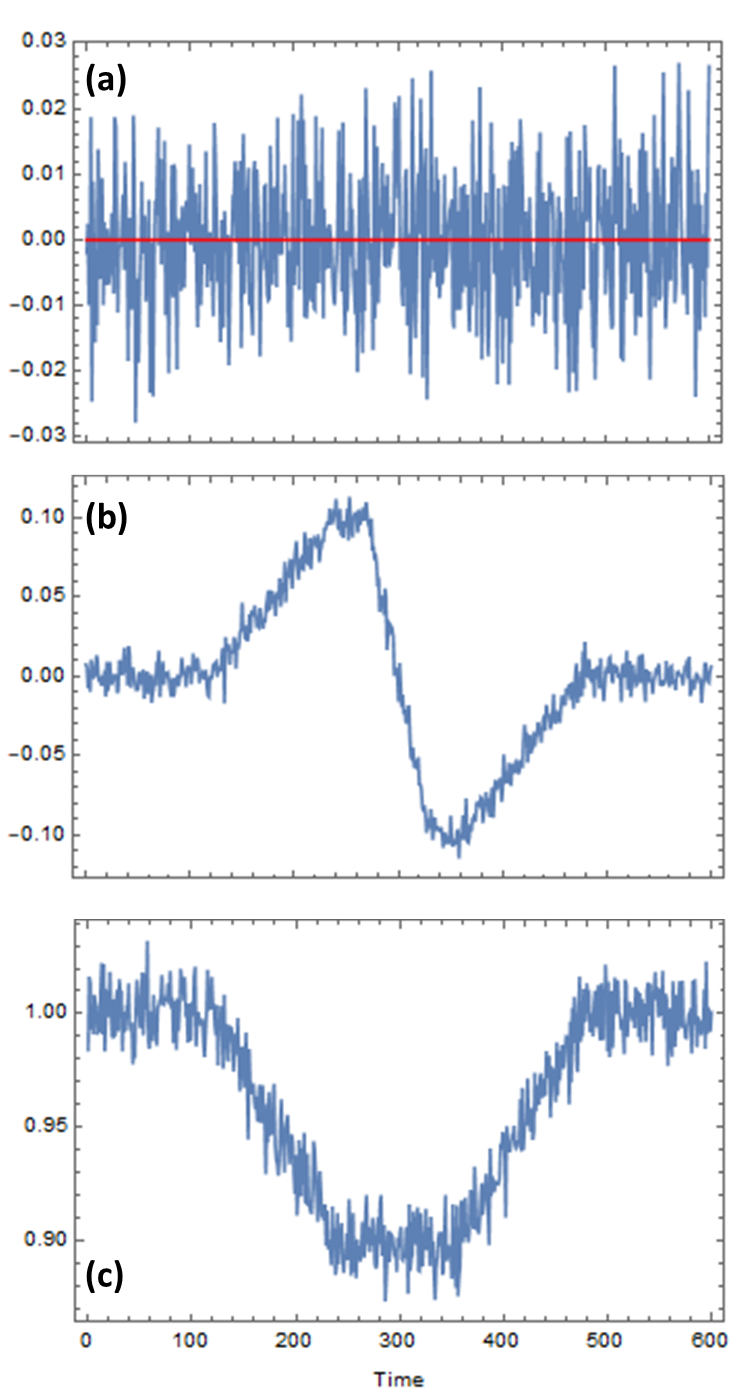}
    \caption{Normalized light curves with astrophysical noise. (a) The Gaussian white noise $\sigma$ added to the background. Here $\sigma_{bg} =0.01$ which means that 68.2\% of the noise falls between -0.01 to +0.01. We varied $\sigma_{bg}$ in our simulations from 0.01 to 0.4. (b) The light curve for the folded transit with the background noise. (c) The light curve for the transit with background noise. The ratio of the radius of the planet to the host star for both (b) and (c) is 1/10.}
    \label{fig:whitenoise}
\end{figure}

The intensity of the host star and background noise are summed up to obtain a signal as a function of time shown in Fig. \ref{fig:whitenoise}(a). For the transit method, shown in Fig. \ref{fig:whitenoise}(c), we see a dip in the intensity, which indicates a planet moving across the host star, while in the case of the folded transit method, the resulting signal of half of the view is deducted from the other half, which results to Fig. \ref{fig:whitenoise}(b).

\begin{figure}[ht]
    \centering
    \includegraphics[width=0.65\linewidth]{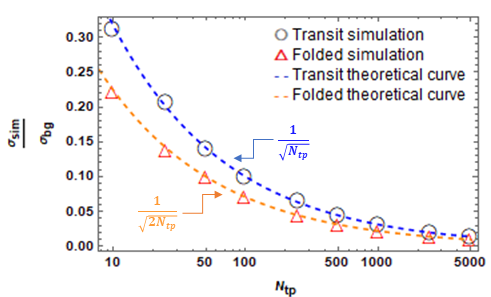}
    \caption{Ratio between the noise from simulations, $\sigma_{sim}$ and the input background noise $\sigma_{bg}$. The folded transit has small ratios than the usual transit because the $N_{tp}$ is technically doubled by the process of folding resulting to a $1/\sqrt 2$ factor between the ratios. This factor is not as important when $N_{tp}$ is large.}
    \label{fig:Ratio}
\end{figure}

It is obvious from Eq. (\ref{eqn:variance}) that the ratio between the $\sigma_{sim}$ and $\sigma_{bg}$ should fall as $1/\sqrt{N_{tp}}$. However, what is not initially intuitive is the lower ratio for the folded transit as shown in Fig. \ref{fig:Ratio}. 

In our simulation, the number of transit observed is the same, but through the process of folding, the number of samples doubles, similar to increasing the exposure time. As such, a factor of $1/\sqrt2$ is seen in the ratio for the folded transit. This has an effect on the $SNR$. 

\begin{figure}[ht]
    \centering
    \includegraphics[width=0.58\linewidth]{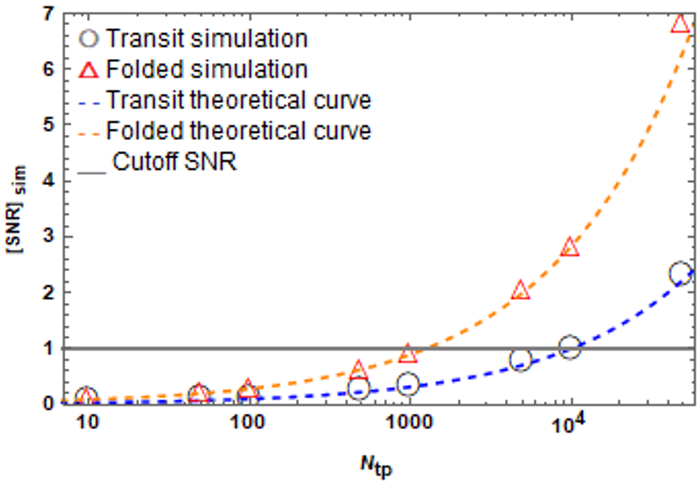}
    \caption{ The $[SNR]_{sim}$ as the $N_{tp}$ is increased. The simulation points sit perfectly with the theoretical curve. The ratio between the $[SNR]_{folded}$ and the $[SNR]_{trans}$ in simulation is similar to the ratio we derived (Eq. (\ref{eqn:SNR_transit}) and Eq. (\ref{eqn:SNR_transitF})), a factor of $2\sqrt{2}$. The horizontal line is at $[SNR]_{sim}=1$.}
    \label{fig:SNR}
\end{figure}

We define the $SNR$ as

\begin{equation}
SNR=\frac{S}{N}=\frac{S}{\frac{\sigma_{bg}}{\sqrt{N_{tp}}}}= \text{const.},
\label{eqn:snr}
\end{equation}

where $S$ is the signal that depends on the planet and the star sizes, the latitude of the transit, and the star's limb-darkening, while $\sigma_{bg}$ is the background noises affecting the instrument's performance.  Equation (\ref{eqn:snr}) is just a rewriting of the general expression for the detection probability in Deeg et al. \cite{Deeg} where we lump the $N_{tp}$ as modifying the $\sigma_{bg}$. By doing so, we show that the noise level for the folded transit is reduced to $1/\sqrt{2}$ compared to the usual transit. The $SNR$ will have an added $\sqrt{2}$ factor regardless of the ratio between the radius of the planet $R_p$ and the host star $R_s$.

The signal in our calculations will come from the ratio of the planet and the star's radius only. We neither consider limb-darkening \cite{Mandel2002} nor the inclination in the transiting planet's motion. The signal for the usual transit is given by the dip in the light curve as,

\begin{equation}
S_{trans}=\frac{\pi R_s^2-\pi R_p^2}{\pi R_s^2}= 1-\frac{R_p^2}{R_s^2}.
\label{eqn:LC_transit}
\end{equation}

The host star's brightness without the planet can be subtracted hence, the signal of interest is $\frac{R^2_p}{R^2_s}$. This gives us an $[SNR]_{trans}$, as in

\begin{equation}
\left[SNR\right]_{trans}=\frac{R_p^2}{R_s^2}\frac{\sqrt{N_{tp}}}{\sigma_{bg}}.
\label{eqn:SNR_transit}
\end{equation}

On the other hand, the signal for the folded transit is given by

\begin{equation}
\begin{split}
S_{folded}&= \left[\frac{\Delta I_2}{I_{total}}\right] - \left[\frac{\Delta I_4}{I_{total}}\right]\\
&=\frac{1}{I_{total}}\left[\Delta I_2-\Delta I_4\right]\\
&=\frac{2\pi R_p^2}{\pi R^2_s-\pi R^2_p}\\
&\approx 2\frac{R_p^2}{R_s^2},
\end{split}
\label{eqn:LC_foldedtransit}
\end{equation}

where the denominator in Eq. (\ref{eqn:LC_foldedtransit}) is expanded and only the leading term is retained. It is obvious that the signal is also doubled in the folded transit. The $[SNR]_{folded}$ is given by

\begin{equation}
\begin{split}
    \left[SNR\right]_{folded}&=2\sqrt{2}\frac{R_p^2}{R_s^2}\frac{\sqrt{N_{tp}}}{\sigma_{bg}}\\
    &=2\sqrt{2}\left[SNR\right]_{trans}.\\
\end{split}
\label{eqn:SNR_transitF}
\end{equation}

The $SNR$ of the folded transit is increased by a factor of $2\sqrt{2}$. The $2$ factor is from the signal, while the $\sqrt{2}$ is from the reduced noise.

Using the same simulation process earlier, we look at the $SNR$ as it changes with $N_{tp}$ using an $R_p/R_s = 1/10$ or a Jupiter-sized planet transiting a Sun-sized host star. We plot this in Fig. \ref{fig:SNR}.

Our simulations confirm our calculations. There are two implications of our results. These are: (1) the folded transit method lessens the number of measurements needed to be done. In Fig. \ref{fig:SNR} for the same $\sigma_{bg}$, an Earth-sized planet transiting a Sun-sized host star can be detected by order of magnitude lesser measurement. If the threshold $SNR$ for detecting this system is 1, meaning the signal is as high as the noise, the folded transit would only need about 1000 measurements compared to the need for 10,000 measurements in the usual transit method; (2) given the same amount of measurement, $N_{tp}$ and background noise $\sigma_{bg}$, the $SNR$ is increased by $2\sqrt{2}$, thereby increasing the detectable limit of the transit method to $1/8^{1/4}$ of the radius of the planet that can be detected by the usual transit method. With this reduction, we will be able to detect sub-Earths (0.8 $R_{earth}$) with the same amount of measurement when detecting super-Earths (1.4 $R_{earth}$) using the transit method. While this might be modest, a reduction is still a welcome enhancement, especially with planets of small radii.
\clearpage
\subsection{Random Sunspots}
\begin{figure}[ht]
    \centering
    \includegraphics[width=0.5\linewidth]{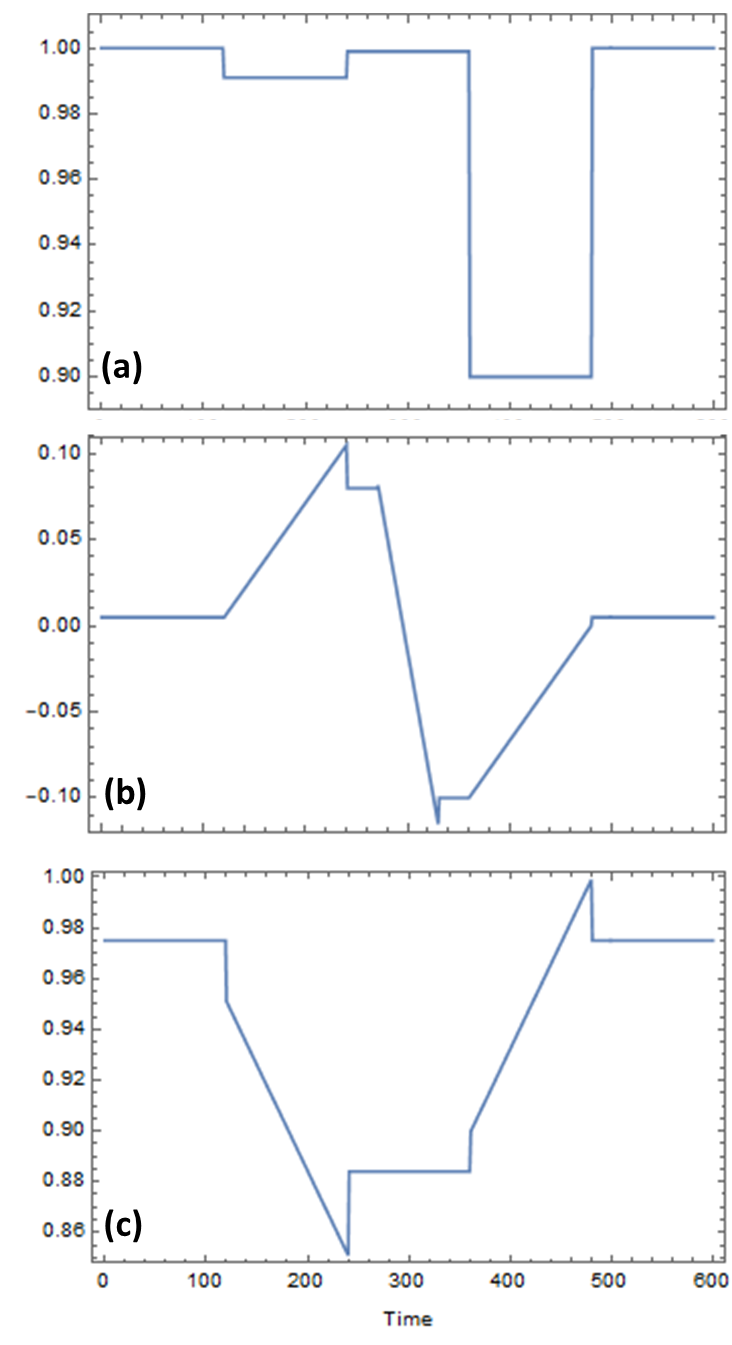}
    \caption{Normalized light curves with sunspots. (a) Sunspots of random size and occurrences were subtracted to the star's intensity. We varied the total area of the sunspots from 1\% to 10\% of the star's area.  (b) The light curve for the folded transit with sunspots. (c) The light curve for the transit with sunspots. The ratio of the radius of the planet to the host star for both (b) and (c) is 1/10.}
    \label{fig:sunspots}
\end{figure}

Similarly, the intensity of a star with random occurrences of sunspots is summed up to produce a signal as a function of time, as shown in Fig. \ref{fig:sunspots}(a). The dips correspond to the existence of a sunspot or group of sunspots in that specific time interval. The intensity drops as the planet moves across the host star for the usual transit method, as shown in Fig. \ref{fig:sunspots}(c) while in the case of the folded transit method, the sum of half of the view is deducted from the other half, which results in Fig. \ref{fig:sunspots}(b).

\begin{figure}[ht]
    \centering
    \includegraphics[width=0.65\linewidth]{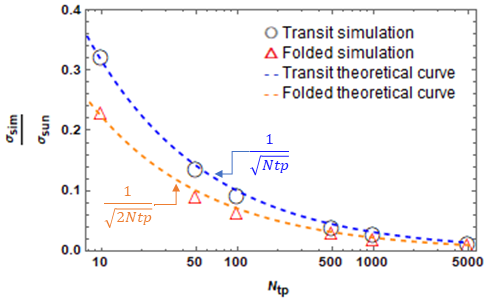}
    \caption{Ratio between the noise from simulations, $\sigma_{sim}$ and the input sunspot noise $\sigma_{sun}$. Again, the folded transit produced smaller ratios than the transit because the $N_{tp}$ is doubled through folding.}
    \label{fig:Ratio1}
\end{figure}

Figure \ref{fig:Ratio1} shows that the ratio between the $\sigma_{sim}$ and $\sigma_{sun}$ follows the curve $1/\sqrt{N_{tp}}$ whereas the ratio for the folded transit falls as $1/\sqrt{2N_{tp}}$. Although the sunspots reduce the intensity much like a transiting planet, the time interval of them appearing is random. Thus, in the limit of many observations, the noise reduces similarly to the Gaussian white noise. 

\begin{figure}[ht]
    \centering
    \includegraphics[width=0.58\linewidth]{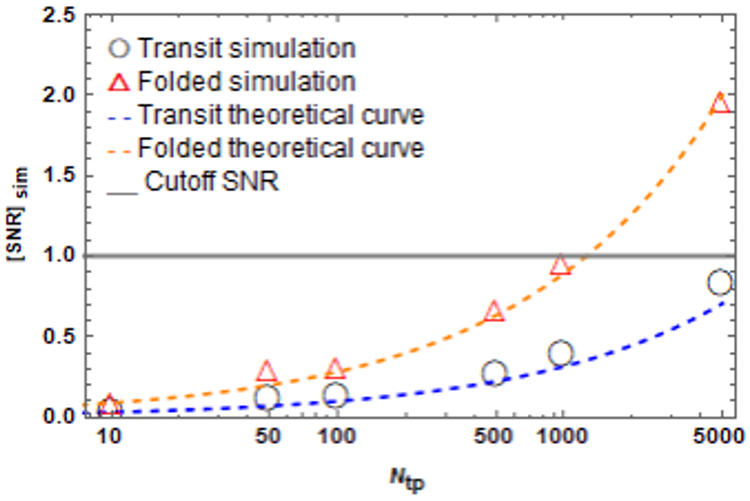}
    \caption{ The $[SNR]_{sim}$ as the $N_{tp}$ is increased. We see that the simulation points consistently sit higher than the theoretical curve in the transit method. This uniform increase signifies an overestimate in the planetary signal.}
    \label{fig:SNR_sunspots}
\end{figure}

Again, we look at the $SNR$ as it changes with $N_{tp}$ using an $R_p/R_s = 1/10$ or a Jupiter-sized planet transiting a sun-sized host star. We  plot this in Fig. \ref{fig:SNR_sunspots}.

The result of our sunspot simulations demonstrates perhaps the most vital characteristic of the folded transit method: it reduces the effect of sunspots in the light curve. With the folded transit, the variability in the peak-to-peak signal is reduced compared to the average in the dip in the usual transit method. This variability reduction directly impacts the measurement of planetary parameters, providing a more accurate estimation. In addition, the reduction can be likened to the centroid detection employed by Bryson et al. \cite{bryson2013identification} to identify false positives in signal from the Kepler data \cite{bryson2013identification}. In this case, the centroid detection is afforded by the folded technique similar to measuring the nanometric displacement of a laser beam (see Hermosa et al.\cite{hermosa2011quadrant} for example). Moreover, if the spot persists similar to the time of transit of the planet, its effect will be a reduction in the total intensity and an uneven peak in the folded transit light curve only. Unfortunately, other common false-positive events such as eclipsing binary and background transiting planet or companion transiting planet may not be discriminated with the folded technique, a problem it shares with the usual transit method \cite{santerne2013astrophysical}. However, methods used to optimize detection and reduce false positives in the usual transit method can also be used in the folded technique \cite{hippke2019optimized,morton2012efficient}.
\clearpage
\section{Execution of a folded detection}
\begin{figure}[ht]
    \centering
    \includegraphics[width=0.85\linewidth]{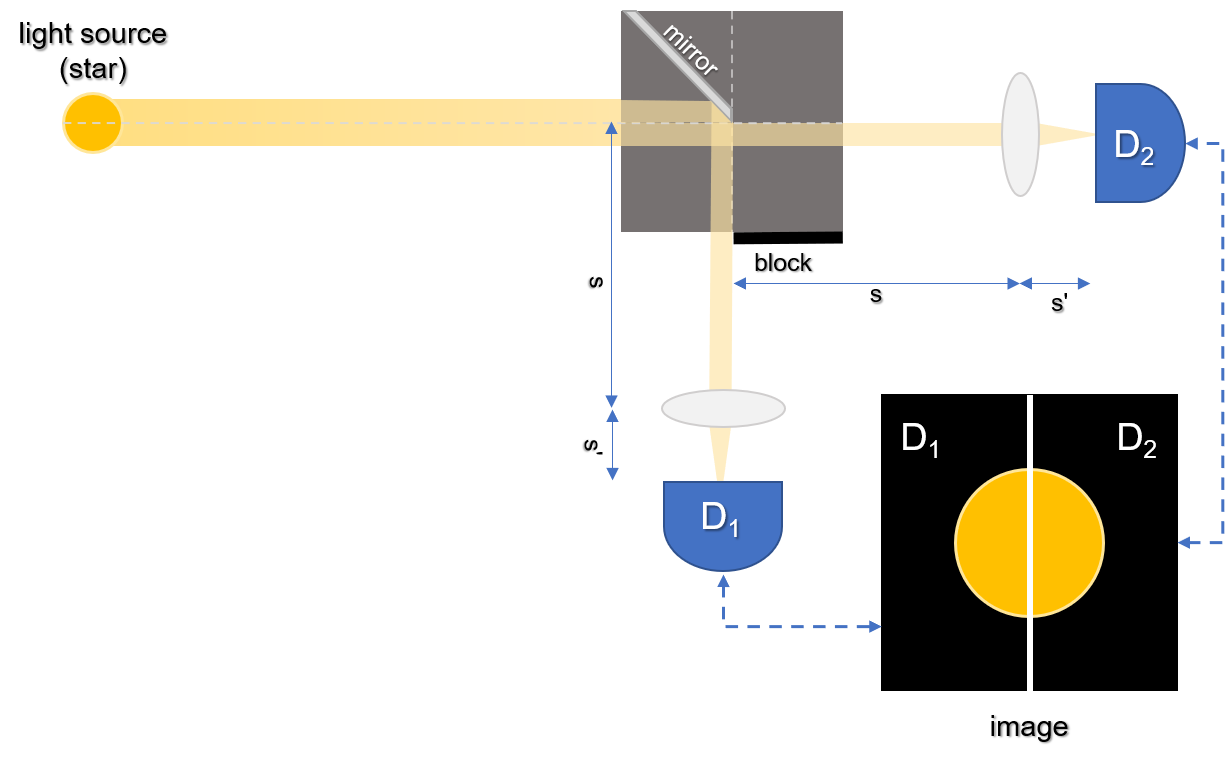}
    \caption{A diagram of a split detector system.} 
    \label{fig:split}
\end{figure}

In folded detection, we will employ the use of a split detector. A split detector, as shown in Fig. \ref{fig:split}, divides the light source into two areas of equal intensities by reflecting half of the intensity distribution to one detector and transmitting the remaining half to another detector. Studies have shown that split detection provides promising results and presents clearer images with increased contrast \cite{split,Sulai:14,cunefare2016automatic}.

\begin{figure}[ht]
    \centering
    \includegraphics[width=0.85\linewidth]{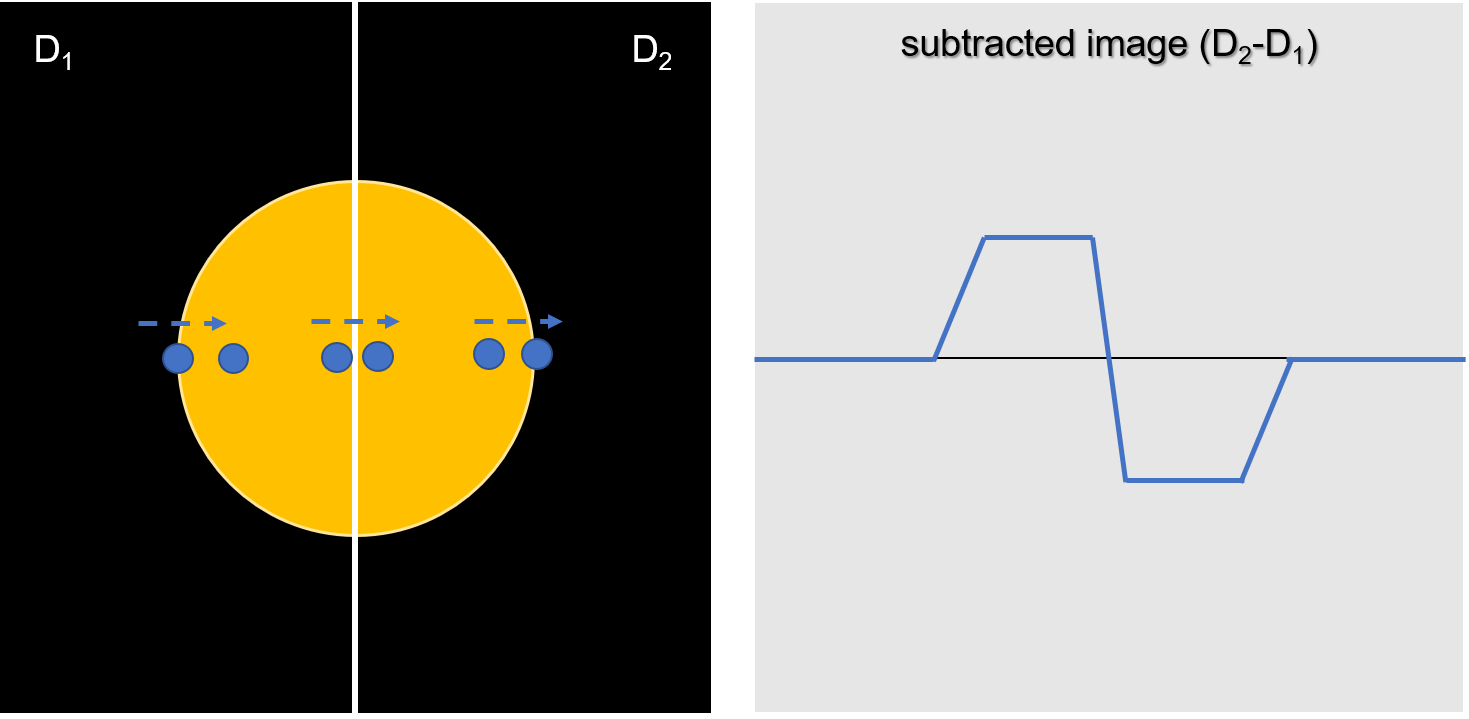}
    \caption{A visual representation of the intensity difference obtained from the two detectors.} 
    \label{fig:split1}
\end{figure}

By subtracting the intensity from the two detectors, we technically remove the host star's signal, our imaging light source, thus isolating the planetary signal and doubling its peak-to-peak amplitude. This is shown in Fig. \ref{fig:split1}.

Essentially, the number of measurements (or transit observations) doubles since the two detectors simultaneously capture a unique measurement and will continue to do so until the planet has completed its transit. 

The folded detection fails if we used the usual detector to capture the transit and directly "fold" the measured transit light curve. If we do that, there will effectively be no gain since the number of data points is conserved.

\newpage

\section{Conclusion}
We created a new technique, the folded transit method, and derived its light curve using simple mathematics. We compared our technique with perhaps the most successful exoplanet detection method, transit photometry. We explored the effects of noises such as background noise and sunspots. We modeled the background noise on a white Gaussian noise process while sunspots were made to behave consistently with their known activity.

In both analytical expression and simulations, we have shown that the folded transit method can increase the signal by a factor of 2 and reduce the effective noise by a factor of $\sqrt{2}$, effectively giving our method an $SNR$ that is a factor of $2\sqrt{2}$ greater than the transit method. This increase enables fewer transit measurements to get a definitive transiting planet signal and detect a smaller planetary radius by as much as $(1/8)^{1/4}$. Further, we have demonstrated that the folded transit method lessens the effect of sunspots in the light curve. This allows a more accurate estimate of planetary parameters.

The motivation for exoplanet hunting is not only to find unmistakable signs of current life but to give us a glimpse of the formation and evolution of planetary systems. Each exoplanet discovered, whether life-sustaining or not, helps us understand the universe more: how it works and how it will potentially be in the future. 
\section{Acknowledgments}
This work was funded by the UP System Enhanced Creative Work and Research Grant (ECWRG 2019-2-06-R).
\newpage
\section*{References}
\bibliographystyle{iopart-num}
\bibliography{ms}

\end{document}